\documentclass[useAMS,usenatbib]{mn2e}
\usepackage{amssymb}
\usepackage{graphicx}
\usepackage{url}
\usepackage{color}
\usepackage{amsmath}

\newcommand{\beq}{\begin{equation}}
\newcommand{\eeq}{\end{equation}}

\newcommand{\apj}{ApJ}
\newcommand{\apjl}{ApJ}
\newcommand{\apjs}{ApJS}
\newcommand{\aap}{A$\&$A}
\newcommand{\aapr}{A$\&$A}

\newcommand{\araa}{ARAA}

\newcommand{\mnras}{MNRAS}

\newcommand{\prd}{PRD}
\newcommand{\physrep}{Phys. Reports}
\newcommand{\aj}{AJ}
\newcommand{\nat}{Nature}

\newcommand\lsim{\mathrel{\rlap{\lower4pt\hbox{\hskip1pt$\sim$}}
        \raise1pt\hbox{$<$}}}
\newcommand\gsim{\mathrel{\rlap{\lower4pt\hbox{\hskip1pt$\sim$}}
        \raise1pt\hbox{$>$}}}

\newcommand{\Msol}{{\rm M}_{\odot}}

\newcommand{\Mpc}{{\rm Mpc}}
\newcommand{\Gpc}{{\rm Gpc}}
\newcommand{\yr}{{\rm yr}}
\newcommand{\Myr}{{\rm Myr}}
\newcommand{\cm}{{\rm cm}}
\newcommand{\erg}{{\rm erg}}
\newcommand{\s}{{\rm s}}
\newcommand{\keV}{{\rm keV}}
\newcommand{\K}{{\rm K}}

\newcommand{\rmd}{{\rm d}}

\newcommand{\Hmol}{{\rm H}_2}

\begin{document}

\title[21~cm signature of SMBH growth]{
The imprint of the cosmic supermassive black hole growth history on the
21~cm background radiation}

\author[T. L. Tanaka, R. M. O'Leary, and R. Perna]{Takamitsu L. Tanaka$^{1,2}$\thanks{E-mail:takamitsu.tanaka@stonybrook.edu},
Ryan M. O'Leary$^{3}$, Rosalba Perna$^{1,3}$\\
$^{1}$Department of Physics and Astronomy, Stony Brook University, Stony Brook, NY 11794, USA\\
$^{2}$Department of Physics, New York University, 4 Washington Place, New York, NY 10003, USA\\
$^{3}$JILA, University of Colorado and NIST, 440 UCB, Boulder, CO 80309-0440, USA
}
\date{}
\maketitle

\label{firstpage}

\begin{abstract}
The redshifted $21~\cm$ transition line of hydrogen
tracks the thermal evolution of the neutral intergalactic medium (IGM)
at ``cosmic dawn,'' during the emergence of
the first luminous astrophysical objects ($\sim 100~\Myr$ after the Big Bang)
but before these objects ionized the IGM ($\sim 400-800~\Myr$ after the Big Bang).
Because X-rays, in particular, are likely to be the chief energy courier for
heating the IGM, measurements of the $21~\cm$ signature
can be used to infer knowledge about the first astrophysical X-ray sources.
Using analytic arguments and a numerical population synthesis algorithm,
we argue that the progenitors of supermassive black holes (SMBHs) should
be the dominant source of hard astrophysical X-rays---and thus the primary driver
of IGM heating and the $21~\cm$ signature---at redshifts $z\ga20$, 
\textit{if} (i) they grow readily from the remnants of Population III
stars and (ii) produce X-rays in quantities comparable to what is observed
from active galactic nuclei and high-mass X-ray binaries.
We show that models satisfying these assumptions dominate
over contributions to IGM heating from stellar populations,
and cause the $21~\cm$ brightness temperature to rise at $z\ga 20$.
An absence of such a signature in the forthcoming observational data would imply
that SMBH formation occurred later (e.g. via so-called direct collapse scenarios),
that it was not a common occurrence in early galaxies and protogalaxies,
or that it produced far fewer X-rays than empirical trends at lower redshifts,
either due to intrinsic dimness (radiative inefficiency)
or Compton-thick obscuration close to the source.
\end{abstract}


\section{Introduction}

Supermassive black holes (SMBHs)
reside in the hearts of most massive galaxies
(see \citealt{KormendyHo13} for a review).
Through luminous quasar episodes, they may play key roles in shaping
their galactic and intergalactic environments \citep[e.g.][]{RicOst04, Cattaneo+09}.
Discoveries of quasars 
at redshifts $z\sim 6-7$ have revealed that SMBHs with masses of several $10^9\,\Msol$
were already in place when the Universe was less than a Gyr old
\citep[e.g.][]{Fan+01, Willott+03,Willott+10a,Mortlock+11,Venemans+13}.

Despite the astrophysical significance of SMBHs and their
presence throughout cosmic time,
their origins are not well constrained by current observations
and remain a subject of active investigation \citep[see reviews by][]{Volonteri10, Haiman13}.

Broadly speaking, theoretical hypotheses of ``seed'' BHs
that grow into SMBHs fall into one of two categories.
The first type of seeds are remnants of the
earliest stars  (Population III, or PopIII, stars),
which form with masses $\sim 10-100\,\Msol$
at redshifts $z\ga20$
\citep{Abel+02,Bromm+02,Yoshida+08, Stacy+10, Greif+11b, Hirano+14},
and subsequently grow through a combination of gas accretion and mergers
(e.g., \citealt{HaimanLoeb01}, \citealt{MadauRees01};
additional references in the detailed description in \S2).
The second class of seeds, called ``direct collapse'' BHs (DCBHs hereafter),
forms with much greater initial masses, $\sim 10^4-10^5\,\Msol$
at later times $z\la 15$.
They form inside gas clouds that do not fragment to form ordinary stars,
but instead collapse to form a much more massive compact object
(e.g. \citealt{BrommLoeb03}, \citealt{Koushiappas+04},
\citealt{Begelman+06}, \citealt{LodatoNatarajan06}).
Both families of seed models require the seed BHs to grow
at rates comparable to the Eddington rate to explain 
the observed $\sim 10^9\,\Msol$ quasar SMBHs at $z\approx 6-7$,
with an average $e$-folding timescale no longer than $ 40-50\,\Myr$ between
seed formation and $z\approx 7$ \citep[e.g.][]{Tanaka14}.

Currently, there are few observational constraints on
the cosmic history of SMBHs at $z>6$.
The $z\sim 6$ quasar luminosity function and inferred underlying SMBH mass function
leaves many degrees of freedom for theoretical explanations.
The main empirical constraints are that models should grow
enough $\sim 10^8-10^9\, \Msol$ SMBHs to account for the quasar observations,
and that they exceed neither empirical estimates of
the universal SMBH mass density nor the unresolved cosmic X-ray background
\citep[e.g.][]{TH09, Salvaterra+12}.
There is no firm empirical evidence that favors either kind of seed model,
places meaningful constraints on how common the seeds were,
or indicates whether they grew steadily (e.g., spurred by the frequency of major
galaxy mergers at these large redshifts \citealt{Li+07, Tanaka14})
or in shorter, intermittent spurts\footnote{
Observations do indicate that the duty cycle of quasars increases
to $\ga 50\%$ at $z>4$, compared to $\la 1\%$ at $z<2$---see, e.g., refs.---suggesting
that SMBH growth is not very intermittent, at least at $4\la z<\la $.
} \citep{VolRees05, Madau+14, Volonteri+15}.
Even future direct observations of quasars at $z\ga7$ may not
be able to distinguish between the PopIII and DCBH seed scenarios---most
published SMBH growth models for both scenarios
are consistent with the existence of $\sim 10^5\,\Msol$ SMBHs at $z\sim 12$
(a condition that comes about naturally if the progenitors of the $z\sim 6-7$,
$\sim 10^9\,\Msol$ quasar SMBHs grew at near the Eddington limit).

In this work, we show that upcoming observations of the sky-averaged redshifted $21~\cm$ line
from the hyperfine transition of neutral hydrogen can help elucidate
the growth history and abundance of SMBHs at $z\ga 20$
\citep{Ricotti+05, Ripamonti+08, Mirocha+13}.
This line appears in absorption if the gas spin temperature is lower than
that of the cosmic microwave background (CMB), and in emission otherwise;
it can thus be used to map the thermal history of the intergalactic medium
prior to cosmic reionization (see \citealt{Furlanetto+06} for a review).
The strength of the 21~cm transition is particularly sensitive to the
thermal state of the gas, due to the coupling of the spin temperature
with the gas temperature via collisions and the Lyman-$\alpha$
background \citep{Wouthuysen52,Field58}.
Astrophysical sources at ``cosmic dawn,'' prior to cosmic reionization ($z \ga 11$),
leave an imprint in this line by building up a Lyman-$\alpha$ background 
and by heating and ionizing the IGM (see, e.g.,
\citealt{Furlanetto06,PritchardLoeb08,Mirocha+13,YajimaKochfar14}).

X-rays can contribute strongly to this signature
because they can raise the IGM spin temperature
to above the CMB temperature before fully reionizing it \citep{RicOst04}.
The imprint should be seen in the sky-averaged (global) signature,
because the long mean-free path of X-rays (with energies $\ga 1~\keV$)
allows them to heat the IGM nearly isotropically.
The two strongest classes of X-ray sources at cosmic dawn
are expected to be seed BHs accreting gas en route to growing into SMBHs
and high-mass X-ray binaries (HMXBs).

In this paper, we argue that if the SMBHs observed as quasars at $z\sim 6$
grew from PopIII seeds via radiatively efficient gas accretion,
they should leave a strong increase in the $21~\cm$ brightness temperature at $z\ga 20$.
The absence of this feature in future observations would
imply either that most SMBHs formed at later epochs,
that their growth was rare,
or that they produce much less energy in X-rays relative to their mass growth
than the standard accretion-disc interpretation of luminous AGN activity.

This paper is organized as follows.
In \S\ref{sec:analytic}, we present a series of analytic estimates
to show that, for a wide range of assumptions,
the X-ray output due to SMBH growth at $z\ga 6$
dominates over that associated with HMXB activity.
\S~\ref{sec:heat} presents the physical and
numerical implementation of the PopIII and the DC models for the
growth of the SMBHs, and the corresponding heating history of the
IGM in the two scenarios. The computation of the 21~cm signal is
presented in \S~\ref{sec:21cm}, together with the contribution from stars.
We summarize our main results and conclude in \S~\ref{sec:summary}.

\section{Analytic Estimates of X-ray Outputs}
\label{sec:analytic}

In this section, we compare the X-ray output from two classes of astrophysical
sources---accreting nuclear BHs in the earliest haloes (galaxies), 
and HMXBs.

\subsection{Estimates of total emitted X-ray energy densities}
The comoving SMBH mass density in the local Universe is $\rho_{\rm SMBH}(z=0)\sim 4\times 10^5~\Msol\Mpc^{-3}$
\citep[e.g.][]{AllerRichstone02, YuTremaine02, Marconi+04, Shankar+04}.
Linking AGN luminosity functions with SMBH mass growth via Soltan's argument \citep{Soltan82},
the same quantity at $z\sim 6$ is estimated to be $\rho_{\rm SMBH}(z\approx 6)\gsim 10^4~\Msol\Mpc^{-3}$, i.e.
ninety per cent is thought to have been accumulated at a radiative efficiency
$\eta \equiv L/(\dot{M}c^2) \sim 0.07$ via gas accretion since $z\sim 6$ \citep{Shankar+09a, Shankar+10}.
On average, AGN emit a fraction $f_{\rm bol}\sim 0.05$ of their total light
in $2-10~\keV$ X-rays \citep[e.g.][and refs. therein]{Hopkins+07b};
accreting stellar-mass BHs and IMBH candidates emit the bulk of their light
in X-rays \citep[e.g.][]{FenderBelloni12},
a property that's consistent with the standard theory
of luminous accretion discs \citep{SS73}, which predicts that lower-mass BHs
have harder accretion spectra at the same Eddington fraction.

Therefore, if the precursors to SMBHs grew via the same accretion mode
as in the standard picture for luminous AGN at lower redshifts,
the total comoving X-ray energy density they would have emitted
prior to $z\sim 6$ can be estimated as
\begin{align}
\epsilon_{\rm BH, X}(z>6)
&\lsim \eta f_{\rm bol}~\rho_{\rm SMBH}(z\approx 6)~c^2 \nonumber\\
&\sim 100 \left(\frac{\eta_{\rm BH}}{0.07}\right)\left(\frac{f_{\rm bol}}{0.05}\right)
~\Msol~c^2~\Mpc^{-3}.
\label{eq:LX_BHs}
\end{align}

Observations show that in the absence of X-ray AGN activity,
star-forming galaxies at low redshift produce, on average,
$2-10~\keV$ X-ray luminosities proportional to their star formation rate $\dot{M}_{\ast}$,
with 
\beq
L_{\rm \ast, X} \sim \ell_{\rm X} \dot{M}_{\ast},
\label{eq:LXSFR}
\eeq
where $\ell_{\rm X} \ga 10^{39}\erg~\s^{-1} \Msol^{-1}\yr $ \citep{Grimm+03, BasuZych+13, Mineo+12, Mineo+14} is the factor of proportionality.
Most of the X-ray luminosity is attributable to HMXB activity, which tracks young stellar populations.
The above relationship can also be expressed as a radiative efficiency
(i.e. the $2-10~\keV$ energy emitted as a fraction of the rest mass energy of stars formed):
\beq
\eta_{\ast,{\rm X}}\equiv\frac{L_{\rm \ast, X} }{\dot{M}_{\ast} c^2} = \ell_{\rm X}c^{-2} \approx 1.8 \times 10^{-8} ~ \ell_{\rm X, 39}, 
\eeq
where $\ell_{\rm X,39}$ is $\ell_{\rm X}$ in units of $10^{39}\erg~\s^{-1} \Msol^{-1}\yr $.

If dark matter haloes with virial temperatures $T_{\rm vir}>10^4~\K$
(in the atomic hydrogen cooling regime) undergo rapid star formation,
and convert $f_{\ast}\sim 10\%$ of their baryonic mass to stars \citep[e.g.][]{FukugitaPeebles04},
then the total  luminosity density of hard X-rays emitted by star-forming galaxies 
prior to $z\sim 6$ is
\begin{eqnarray}
\epsilon_{\rm \ast, X}(z>6) &\sim& \eta_{\ast,{\rm X}}f_{\ast}
\frac{\Omega_{\rm b}}{\Omega_0}\rho_{\rm halo}(M>M_4; ~ z=6)
\nonumber\\
&\sim& 0.8~ \ell_{\rm X,39}\left(\frac{f_{\ast}}{0.1}\right) ~\Msol~c^2~\Mpc^{-3}.
\label{eq:LX_stars}
\end{eqnarray}
Above, $M_4\equiv M(T_{\rm vir}=10^4~\K)$ is the atomic-cooling halo mass threshold,
$\Omega_{\rm b}/\Omega_0\approx 0.15$ is the ratio of the baryonic fraction to the matter fraction
and $\rho_{\rm halo}(M>M_4;~z=6)\approx 3\times 10^9~\Msol~\Mpc^{-3}$ is the $z\sim 6$ comoving
mass density locked inside dark matter haloes above this threshold
(computed using the Sheth-Tormen mass function,  \citealt{ShethTormen02}).
Note that this estimate for $\epsilon_{\rm \ast, X}$ is dominated by stars in low-mass haloes just above
the atomic-cooling threshold $M_4$.

The ratio of total X-ray energy density emitted by SMBH progenitors
to that emitted by HMXBs is then
\beq
\frac{\epsilon_{\rm BH, X}(z>6) }{\epsilon_{\rm \ast, X}(z>6)}
\sim 100 \left(\frac{\eta_{\rm BH}}{0.07}\right)\left(\frac{f_{\rm bol}}{0.05}\right) 
\ell_{\rm X,39}^{-1}\left(\frac{f_{\ast}}{0.1}\right)^{-1}.
\label{eq:lumXcomp}
\eeq

It's plausible that the earliest galaxies emitted more X-rays per unit mass of star formation---i.e.
that they had more prolific HMXB activity and thus a systematically higher value of $\ell_{\rm X}$.
\cite{BasuZych+13} concluded that this quantity is consistent with rising as $\propto (1+z)$ out to $z\approx 4$.
Using the unresolved cosmic X-ray background as a constraint,
\cite{Dijkstra+12} ruled out $\ell_{\rm X}(z)$ that rises more steeply than
a power-law $\propto (1+z)$ for $\ell_{\rm X,39}(z=0)=3$.
Several studies \citep{Fragos+13a, Fragos+13b, Hummel+14, Ryu+15}
have suggested that $\ell_{\rm X}$ may be a factor $\sim 1-100$ higher than
what is observed in lower-$z$ galaxies.
Even accounting for the possibility of higher $\ell_{\rm X}$ for the earliest stellar populations,
equation (\ref{eq:lumXcomp})  suggests that $\epsilon_{\rm BH, X}(z>6) \gg \epsilon_{\rm \ast, X}(z>6)$---at $z>6$,
and that total X-ray production from SMBH growth should have dominated over that from stellar populations.

We should also note that the above estimates for the SMBH mass density
only account for the massive nuclear BHs identified in low-redshift observations.
For example, it is possible that intermediate-mass ($<10^5\Msol$)
reside in the outskirts of massive galaxies \citep[e.g.][]{Islam+03, OLearyLoeb09, Micic+11, RashkovMadau14}
or in the nuclei of dwarf galaxies \citep[e.g.][]{Ho+97, Izotov+08, Barth+04, Reines+13}.
The expressions above do not account for the $z>6$ X-ray emission from the
growth of such BHs.

Note that the above estimate applies to both PopIII and DCBH
seed scenarios, as long as they arrive at the $z\approx 6$ comoving SMBH mass density
of $\sim 10^4~\Msol ~\Mpc^{-3}$ via luminous gas accretion.

\subsection{Estimates of X-ray luminosity densities }
In a similar vein, we can also estimate the luminosity densities
(as opposed to the cumulative emitted energy densities)
at a given redshift as a function of unknown physical parameters.
This quantity is more relevant for predicting the X-ray background as
a function of redshift.

Suppose that central BHs in galaxies possess an average fraction $f_{\rm BH}$ of the dark matter halo mass,
and that at any given time they shine at an average fraction $f_{\rm Edd}$
(allowing that a fraction of BHs are inactive, and that not all haloes/galaxies
host a central BH).
Then the $2-10~\keV$ luminosity density of active BHs can be written as
\begin{eqnarray}
l_{\rm BH, X} &=& 1.3\times 10^{38} f_{\rm bol} f_{\rm BH} f_{\rm Edd} \nonumber\\
&& \qquad \times ~ ~ \rho_{\rm halo} (M>M_{\rm host})
~\erg~\s^{-1} \Msol^{-1},
\end{eqnarray}
where 
$1.3\times 10^{38} \erg~\s^{-1} \Msol^{-1}$ is the ratio of the Eddington luminosity
of an object to its mass,
$M_{\rm host}$ is the minimum characteristic mass for a dark matter halo
to host a nuclear BH
(i.e. the halo mass scale above which the fraction of haloes that
host a BH is close to one),
and $\rho_{\rm halo} (M>M_{\rm host})$ is the universal density of mass
locked in DM haloes above that mass.

The average Eddington ratio $f_{\rm Edd}$ (over all haloes with $M>M_{\rm host}$,
including those that do not contain an accreting nuclear BH)
may be much higher at $z\ga 6$ than in the local Universe, where it is $f_{\rm Edd}(z=0)\ll 1$.
Clustering data suggest that the duty cycle of quasars increases toward high redshift,
reaching $\sim 0.5$ at $z\approx 5$ \citep[e.g.][]{Shankar+09a},
compared to $\ll 1$ at $z\la 2$.
Such prolific activity of high-redshift SMBHs could be explained
if their growth is triggered by the major merger rate of galaxies \citep{Li+07, Tanaka14},
which scales as $\sim (1+z)^{5/2}$.

Turning to the X-ray luminosity from HMXBs, we can write equation (\ref{eq:LX_stars}) as a luminosity
density that scales with the universal star formation rate density,
\begin{eqnarray}
l_{\rm \ast, X}
&=& \ell_{\rm X} ~\frac{\rm d}{{\rm d}t} ~\rho_{\rm halo}(M>M_4)\nonumber\\
&=& \ell_{\rm X} ~ f_{\ast}\frac{\Omega_{\rm b}}{\Omega_0}~\left|\frac{{\rm d}z}{{\rm d}t}\right|
\zeta(z) \rho_{\rm halo}(M>M_4)\nonumber\\
&\approx& 5.7\times 10^{28} \ell_{\rm X,39} \left(\frac{f_{\ast}}{0.1}\right) 
\left(\frac{1+z}{21}\right) ^{5/2} \zeta(z) \nonumber\\
&~& \qquad \times ~~ \rho_{\rm halo}(M>M_4) ~\erg~\s^{-1} \Msol^{-1}.
\end{eqnarray}
Above, $\zeta(z) \equiv -{\rm d} \ln \rho_{\rm halo}(M>M_4)/{\rm d}z$
is a dimensionless quantity that is of order unity at redshift values of interest in this paper.

The ratio of the X-ray luminosity density produced by accreting BH seeds
to that produced by HMXBs is:
\begin{eqnarray}
\frac{l_{\rm BH, X}}{l_{\rm \ast, X}}
&\sim& 5\times 10^3
\ell_{\rm X,39}^{-1}
f_{\rm Edd}
\left(\frac{f_{\rm BH}}{10^{-3}}\right) 
\left(\frac{f_{\rm bol}}{0.05}\right) 
\left(\frac{f_{\ast}}{0.1}\right) ^{-1}\nonumber\\
&~&~\times
\left(\frac{M_{\rm host}}{M_4}\right)^{-1}
\left(\frac{1+z}{21}\right) ^{-5/2}
\zeta^{-1}(z).
\end{eqnarray}
Above, we have used the fact that the DM halo mass function
${\rm d}n/{\rm d}M_{\rm halo}$ roughly scales as $M_{\rm halo}^{-2}$
at masses below the exponential cutoff of the mass function.
Once again, we arrive at the conclusion that X-ray emission from growing seed BHs
very plausibly dominated over the X-ray emission from HMXBs,
\textit{unless} growing nuclear BHs were rare compared to star-forming galaxies,
or these BHs emitted much less energy in X-rays per unit mass growth than their present-day counterparts.

It is important to note that the above estimates hold even if only a small
fraction of the first PopIII-forming minihaloes formed seed BHs
with $\sim 100~\Msol$.
This is because such haloes merge rapidly, and the fraction of haloes 
containing a seed BH rapidly approaches unity (see e.g. \citealt{TH09}).
For example, a typical halo with a mass $>10^{12}~\Msol$ at $z\approx 6$
will have had hundreds of thousands to millions of progenitor haloes that formed PopIII stars.
Below, we present results from one model that assumes that 
a X-ray luminous PopIII seed BH forms in all 
haloes that reach a virial temperature of $2000~\K$,
and another model that assumes that they form in only $1\%$ of such haloes.
Both models arrive at approximately the same SMBH population by $z\approx 6$
\citep[cf.][]{TH09, TPH12}.

\section{Theoretical model}
\label{sec:heat}

We summarize our computational scheme as follows.
\begin{enumerate}
\item Using a Monte Carlo merger tree code, we simulate the assembly history of
dark matter haloes from $z\sim 50$ to $z=6$. The algorithm \citep{Zhang+08} and halo sample,
which reproduce the \cite{ShethTormen02} mass function with high fidelity,
are the same as those described in \cite{TL14}.
We also account for the suppression of PopIII seed formation
at high redshifts due to supersonic coherent motions of baryons
against dark matter \cite[][and refs. therein]{TLH13}.
\item We plant seed BHs in DM haloes that satisfy specific analytic criteria,
motivated by the physical models considered. In  models with PopIII seed BHs,
the seeds are allowed to form when the halo reaches a virial temperature of $2000~\K$,
which is the approximate threshold for PopIII star formation.
In models with DCBH seeds, we plant central BHs with masses of $10^4$-$10^5\,\Msol$
in a small fraction of haloes above the atomic-cooling threshold.
\item We allow BHs to grow via gas accretion, again following semi-analytic prescriptions
following plausible BH-halo scaling relations (motivated by empirical BH-galaxy scaling relations),
while also satisfying observational constraints such as the universal SMBH mass density.
In our fiducial prescription, BHs accrete to approach a relation $M_{\rm BH}(M_{\rm halo},z)$
motivated by \cite{Ferrarese02}, at accretion rates capped at the Eddingtom limit.
While we treat the accretion as being continuous, quantitatively similar SMBH populations
emerge in models where BH growth occurs sporadically, e.g. triggered by major mergers of the host DM halo \citep{TPH12, Tanaka14}.
We model the X-ray emission of BHs using standard accretion flow theory, and assume 5 per cent
of the emitted energy is reprocessed into a power-law corona with $E>1~\keV$ via Compton upscattering \citep[e.g.][]{Hopkins+07b, Done+12}.
In all models considered in this paper, BH growth and HMXB activity occur only
when the host halo is atomic-cooling ($T_{\rm vir}>10^4~{\rm K}$).
\item We allow the central BHs to merge after their host haloes merge
(as long as the halo merger timescale does not exceed the Hubble time, in which case
the haloes are assumed to become satellites),
and determine semi-analytically whether the merged products
are ejected or retained after undergoing gravitational recoil \citep[see][for details]{TH09}.
\item Concurrently with the BH growth, we follow the universal star-formation rate
and the corresponding X-ray emission from young stellar populations,
by assuming that the star formation rate scales with the increase in the baryonic
mass content of the halo (see the previous and following sections).
\item At each timestep in the merger tree code, we track the cumulative X-ray background
for photons with sufficiently long mean-free-paths ($\ga 1~\keV$) to form an isotropic
background.
We account for absorption through a neutral IGM, as well as redshifting.
We use the time-dependent X-ray background to compute
the evolution of the mean IGM kinetic temperature $T_{\rm IGM}(z)$,
which in turn is used to give the global (sky-averaged) 21 cm spin temperature.
We follow \cite{PritchardFurlanetto06} and \cite{FurlanettoPritchard06} in
evaluating the Ly$\alpha$ background and UV emission from stellar populations.
\end{enumerate}

Most of the key model ingredients described above,
such as the X-ray output from the first galaxies
and the growth of seed BHs, contain large theoretical
uncertainties (and are active topics of research in their own right).
The goal of this work is not to make specific predictions for every combination
of plausible theoretical models,
but rather to illustrate the fact that different families of models can correspond to
broadly quantitatively distinct predictions for the global 21 cm signature.
Whereas previous analyses assumed that the growth of the X-ray background
scales directly with the growth in the overall mass density locked inside
collapsed dark matter haloes, here we account for relevant effects
in the hierarchical SMBH-halo assembly process, such as
individually limiting BH growth to the Eddington limit, 
occupation fractions and BH-halo scaling relations,
as well as BH mergers and the associated gravitational recoil effect.
\subsection{Global signatures of X-ray emission on the 21cm transition line}
\label{sec:21cm}

Prior to reionization, the thermal history of the IGM is most directly observable using the highly redshifted $21\,$cm
line of neutral hydrogen. The line is globally observable in contrast to the temperature of the CMB with a brightness temperature of 
\begin{equation}
T_b^{21\,{\rm cm}} = 41.5 x_{\rm H} \left(1 - \frac{T_{\rm CMB}(z)}{T_{21}}\right) \left(\frac{1+z}{21}\right)^{1/2}\,{\rm mK},
\end{equation}
where $x_H$ is the neutral hydrogen fraction, $T_{\rm CMB}(z)$ is the CMB temperature at redshift $z$, and $T_{\rm 21}$ is the spin temperature of the neutral hydrogen \citep[see, e.g.,][for a more thorough discussion]{Furlanetto06}.  The 21\,cm spin temperature of HI  couples to the thermal temperature of the gas mainly through the scattering of Lyman-$\alpha$ photons \citep{Wouthuysen52,Field58} from the first stars, with
\begin{equation}
T_{21}^{-1} = \frac{T_{\rm CMB}^{-1}+T_{\rm IGM}^{-1}(x_\alpha+x_c)}{1+x_\alpha+x_c},
\end{equation}
where $x_\alpha$ is the coupling coefficient of the 21~cm line to Lyman-$\alpha$
radiation and $x_c$ is the collisional coupling coefficient, whose contribution is negligible at $z \lesssim 40$.
Once the star formation rate grows to be $\gsim 10^{-3} M_\odot\,$yr$^{-1}$, there are a sufficient number of Lyman-$\alpha$
photons to fully couple the spin temperature to the thermal
temperature of the IGM, $T_{21} \approx T_{\rm IGM}$ \citep{McQuinnOLeary12}.

We compute the coefficient $x_\alpha$ following \cite{PritchardFurlanetto06},
summing the flux across the Lyman series weighted by ``recycling coefficients.''
The Lyman-$\alpha$ background is computed by
assuming that 9690 photons are produced, per baryon,
between Lyman-$\alpha$ and the Lyman limit,
with a power-law spectrum $\varepsilon_\nu\propto \nu^{-0.86}$
(the PopII star spectral model of \citealt{BarkanaLoeb05}).
We also calculate the dimensionless factor $S_\alpha$ that goes into computing
the coefficient $x_\alpha$, following the wing approximation results of
\citeauthor{FurlanettoPritchard06} (\citeyear{FurlanettoPritchard06}; instead of assuming $S_\alpha=1$).

Generally, the kinetic temperature of the IGM evolves according to
\begin{equation}
\label{eq:heatrate}
\frac{\rmd u}{\rmd t} = -p \frac{\rmd}{\rmd t}\left(\frac{1}{\rho}\right)+\frac{\lambda_{\rm net}}{\rho},
\end{equation}
where $u$, $p$ and $\rho$ are the mean specific internal energy, pressure and density of the IGM.
The first term on the right-hand side, involving the derivative of the IGM density,
is dominated by adiabatic cooling due to cosmic expansion.
The second term includes line and continuum cooling, as well as Compton heating/cooling.
It also includes radiative heating by X-rays, which is the focus of this work.
In this work, we calculate radiative transfer and chemistry for the
species H, H$^+$, He, He$^+$ and He$^{++}$.
We refer the reader to \cite{TPH12} for further details.

In a neutral IGM,  photons with energies above $\approx 1 \,{\rm keV}$
have mean free paths longer than the Hubble horizon,
and will build up an X-ray background that isotropically heats the IGM. 
We assume that the primary astrophysical sources of X-rays are
young stellar populations and  (mini-) quasars, i.e.
SMBHs (or their progenitors) that are undergoing luminous accretion.
Below, we describe our theoretical treatment of each of type of X-ray source.

\subsection{X-ray Sources}
\subsubsection{Stellar populations (HMXBs)}

Following \cite{Furlanetto06}, we assume that the global  $>1~\keV$ X-ray emission
from high-redshift galaxies is proportional to the star formation rate
(as is observed for local star-forming galaxies; see \S2),
and that as DM haloes grow in mass, they convert a fraction
$f_{\ast}\sim 0.1$ of their baryons into stars.
\footnote{Although we consider HMXBs as the chief emitter of X-rays from stellar populations,
other sources, such as supernova remnants, could also contribute.
However, such sources can be incorporated into the quantity $\ell_{\rm X}$
(X-ray emission per star formation rate), as long as they are associated
with young stars.}

Following eq. \ref{eq:LXSFR}, the luminosity density in $2-10~\keV$ X-rays
produced by star formation in the early Universe at any redshift can be estimated as
\beq
\label{eq:epssf}
\epsilon_{\rm 2-10~keV} = \ell_{\rm X,\ast}  \rho_{\ast} = \ell_{\rm X,\ast} f_{\ast} \frac{{\rm d}}{{\rm d}t}\rho_{\rm halo}(M>M_4), 
\eeq
where $\rho_{\rm halo}(M>M_4)$ is the density of matter locked inside haloes above the atomic-cooling threshold.
We assume that the spectral energy distribution from young stellar populations 
takes the form of a power-law
in the relevant energy range (i.e. hard enough to form a background),
with index $\Gamma=1.8$ \citep[e.g.][]{Swartz+04}; equation (\ref{eq:epssf}) gives the normalization to this power-law.

As stated in \ref{sec:heat}, we only allow HMXBs and seed BH growth in haloes
with $T_{\rm vir} > 10^4~\K$. This is a conservative prescription, based on the
simple fact that ionization resulting from mini-quasar activity can unbind
the gas from haloes with shallower potentials
\citep[see also][for additional possible consequences of radiative feedback]{Alvarez+09, Milos+09, ParkRicotti12}.

In principle, haloes in the range $2000~\K \la T_{\rm vir}\la 10^4~\K$
could also produce X-rays via nuclear BH growth or HMXB activity.
Such a scenario would result in additional heating as early as $z\approx 30$,
with a corresponding turn in the 21 cm signature.

It is uncertain whether the quantity $\ell_{\rm X,\ast}$, which has been measured to be
$\ga 10^{39}\erg\s^{-1}\Msol^{-1}\yr$ in star-forming galaxies in the local Universe \citep[e.g.][]{GloverBrand03, Mineo+12}, 
evolves with redshift. 
Following \cite{Dijkstra+12}, we parameterize this uncertainty by adopting a functional form
\beq
\ell_{\rm X,\ast}(z) =\ell_{\rm X,\ast}(z=0)\times (1+z)^b.
\eeq
We adopt $\ell_{\rm X,\ast}(z=0)=3\times 10^{39}\erg\s^{-1}\Msol^{-1}\yr$,
and consider two extreme cases for the parameter $b$:
(i) a conservative case $b=0$ (i.e. no redshift evolution),
and (ii) a more prolific case $b=1$, which is consistent with observations out to $z\approx 4$ \citep{BasuZych+13},
but lies near the $1-\sigma$ exclusion limit for higher redshifts \citep{Dijkstra+12}.

\subsubsection{Seed BHs}

We consider both PopIII and DCBH seed models.

For the PopIII case, we assume that seed BHs can form
once a dark matter halo reaches a virial temperature of $2000~\K$,
which is roughly the threshold required for molecular hydrogen to form,
leading to PopIII star formation.
Once a DM halo reaches this virial temperature, a seed BH forms in
the mass range $10~\Msol < M < 300 \Msol$ with a mass function and power-law
slope $-1.35$, excluding the pair-instability window of $140-260~\Msol$.
While the mass distribution of PopIII stars remains a topic of active research \citep{Hirano+14,Hirano+15},
we believe our choice of a steep power-law is conservative as only a small fraction
of haloes form seed BHs with masses $\ga 100~\Msol$.
We consider a case where a PopIII seed BH forms in each halo reaching $T_{\rm vir}=2000~\K$,
and another in which a seed BH forms in only $1$ per cent of such haloes.

For the DCBH case, we assume that a seed BH with a mass of $10^5~\Msol$
forms in a small fraction of haloes reaching the atomic-cooling threshold of $T_{\rm vir}=10^4~\K$.

This corresponds to a scenario in which pristine atomic-cooling haloes
are irradiated by UV light from nearby star-forming galaxies,
thereby suppressing ordinary PopIII star formation via photodissociation
of $\Hmol$ molecules. Under such conditions, the gas can only cool via atomic
cooling transitions, and collapses isothermally at a temperature $\sim 10^4~\K$,
forming a so-called ``supermassive'' ($\sim 10^5~\Msol$) star that leaves behind
a BH of similar mass via general-relativistic instability.
While the exact conditions necessary for this mode of BH formation,
and how often they occur in nature, remain uncertain, recent theoretical estimates
suggest that DCBH formation may be able to explain the abundance
of the most luminous $z\sim 6$ quasar SMBHs
(comoving number density of $\sim 1~ \Gpc^{-3}$ at $z\approx 6$)
but not the general SMBH population \citep{Dijkstra+14, Latif+15, InayoshiTanaka14}.

We consider a somewhat generous DCBH-seeding prescription,
in which $0.01$ per cent ($10^{-4}$) of all haloes reaching the atomic-cooling threshold
within a redshift range $20>z>10$ form a DCBH of mass $10^5~\Msol$.
This seeding prescription is sufficient to ensure that most galaxy-class haloes
host a SMBH by $z\approx 6$ and at later epochs.

\subsection{Limitations}

In this work, we focus on the initial turnover feature of the global 21 cm signature.
We do not follow the signature past $z\approx 12$, for the following two reasons.

First, our computational algorithm tracks the mean thermal evolution
of the IGM due to an isotropic X-ray background.
Because the code does not track the spatial distribution of the dark matter haloes,
we cannot follow the localized, expanding ionized regions that form around
individual star-forming galaxies and quasars due to their UV emission (whereas
$>1~\keV$ X-rays propagate for  $\sim 1~\Gpc$ before being absorbed).
The strength of the 21 cm signature depends on the column
density of neutral hydrogen in the IGM; because our computational
method cannot reliably track the latter quantity in an anisotropically ionized Universe,
we refrain from modeling the signature into the epoch of UV-driven reionization, i.e. $z\lsim 12$.

Second, it is plausible that the heating of the IGM
affects star formation and (mini-) quasar activity in low-mass galaxies
\citep{Ripamonti+08, TPH12}.
Once the IGM temperature rises,
gas infall into galaxies that fall below the Jeans collapse scale
(or the filtering mass scale; e.g. \citealt{Gnedin00, NaozBark07})
may be suppressed, quenching star formation and (mini-) quasar
activity in these sites.
As a reference, the Jeans mass scale becomes larger than
the atomic-cooling halo virial mass threshold at $T_{\rm IGM}\sim 10^3~\K$.
Because there are numerous uncertainties regarding the details of this global,
thermally driven feedback, we do not attempt to model the regime
where it becomes important.

In other words, the initial rise in the global 21 cm signature at $20\la z \la30$
depends only on the heating rate of the IGM during this epoch,
whereas the signature at later times depends on
the progress of reionization, as well as how effectively
gas from a heated IGM can collapse into low-mass haloes.
We concentrate on the former signature.

\section{Results}

\subsection{IGM kinetic temperature}
\label{sec:IGM}
\begin{figure}[t]
\begin{center}
\includegraphics[width=\columnwidth]{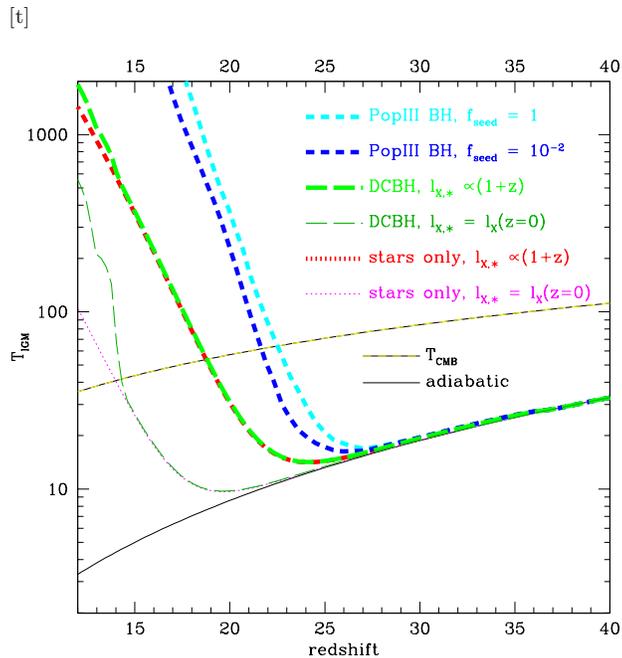}
\end{center}
\caption{The IGM temperature (as contrasted to pure adiabatic cooling)
  resulting from different sources of heating: direct collapse BHs (green and dark green lines) and
  BHs from PopIII seeds (blue and cyan lines), both including the contribution from
  HMXBs. The effect of HMXBs alone is also displayed separately (magenta and red lines).
  Two prescriptions (compatible with independent constraints) are
  considered for each heating source.  Heating from PopIII BHs
  dominates over that of HMXBs alone, while the contribution from
  DCBHs is comparable or subdominant with respect to that of HMXBs,
  until $z\approx 15$.
  For reference, we have also plotted what the IGM temperature would
  be with adiabatic cooling alone (solid black curve),
  as well as the CMB temperature (which determines when the
  21 cm line goes into emission; yellow-and-black dashed curve).
 \label{fig:TIGM}}
\vspace{-1\baselineskip}
\end{figure}

In Figure~\ref{fig:TIGM}, we show the redshift evolution of the mean IGM
kinetic temperature. We will describe the curves in order, roughly from bottom to top, as follows.
The solid black line at the very bottom shows, for reference,  the null case with adiabatic
cooling due to cosmic expansion and no heat sources.

With the exception of the adiabatic case, all of the models include
X-ray heating from HMXBs.
In models with thin lines, it's assumed that the factor
$\ell_{\rm X,\ast}$ (which relates X-ray production to star formation rate)
is the same as measured in star-forming galaxies the local-Universe;
in those with thick lines, this quantity increases toward high redshift as $\ell_{\rm X,\ast}\propto (1+z)$.

The thin, magenta line and the thick, red line (both dotted) show cases
where HMXBs are the only X-ray sources heating the IGM.
The former, which assumes that X-ray production per star formation rate (the quantity $\ell_{\rm X,\ast}$)
does not evolve with redshift, results in the IGM kinetic temperature increasing at $z\sim 20$.
For the latter case, which assumes that $\ell_{\rm X,\ast}\propto (1+z)$,
$T_{\rm IGM}$ increases at $z\sim 25$.

The thick, green curve and the thin, dark green curve (both long-dashed) show models where DCBHs
with $M=10^5\Msol$ form in $0.01\%$ of atomic-cooling haloes at $z<15$,
and grow to match the prescribed relationship $M_{\rm BH}(M_{\rm halo}, z)$, with accretion capped at the Eddington rate.
Note that this seeding fraction of $0.01\%$ is much larger than found in recent theoretical work,
which suggest that DCBHs may have challenges in matching the observed abundance
of high-redshift quasars \citep{Dijkstra+14, Latif+15, InayoshiTanaka14}.
These heating curves are the same as HMXB-only cases described in the previous paragraph,
except for the additional heating at late times ($z\la 15$) due to the emergence
of DCBH seeds.

Finally, the blue and cyan lines (both thick and short-dashed)
show cases in which a PopIII seed BH forms
in haloes reaching virial temperatures $T_{\rm vir}=2000~\K$,
and promptly begins to accrete gas as the host halo also grows.
The cyan line shows the case where such a growing seed forms in all $T_{\rm vir}>2000~\K$
haloes, and the blue line shows the case where a seed forms in $1$ per cent of such haloes.
Both cases assume an HMXB contribution $\ell_{\rm X,\ast}\propto (1+z)$,
but it is clear that the mini-quasar activity of PopIII BHs dominate.

The PopIII seed models heat the IGM much earlier than models
that include HMXB heating only (which, again, use the growth in $\rho_{\rm halo}(M>M_4)$
as a proxy for star-formation; \citealt{Furlanetto06}).
Again, the former assumes that seed BHs form in $>1\%$ of $T_{\rm vir}\approx 2000~\K$
haloes, and grow to match extrapolated BH-halo scaling relations in all haloes with $T_{\rm vir}>10^4~\K$.

It is important to point out that even if PopIII remnants produce copious X-rays,
they may not contribute significantly to the present-day X-ray background.
This is because their spectra are expected to be relatively soft,
as well as sharply peaked in intrinsic X-rays.
For example, a blackbody, Eddington-rate accretion disc around a $10^3~\Msol$ BH
has a spectrum that peaks at $\approx 0.3~\keV$ and drops off by orders of
magnitude above $\approx 1~\keV$.
A graybody disc \citep{Blaes04, TM10} can be harder and peak at $\approx 1~\keV$.
In either case, the bulk of the emission would be redshifted to energies well below
the $\approx 0.1~\keV$ limit of soft X-ray observatories.
Therefore, it is plausible that the vast majority of the X-rays emitted by miniquasars at $z\gsim 20$
is redshifted into the FUV, and be minimally reflected in the present-day X-ray background.
Note the contrast to present-day AGN, whose intrinsic spectra peak
in the UV and whose X-ray emission consists of a power-law tail
(cf. \citealt{Salvaterra+12}, who consider limits on early SMBH
growth by assuming power-law X-ray spectra).

\subsection{21~cm brightness temperature}

\begin{figure}[t]
\begin{center}
\includegraphics[width=\columnwidth]{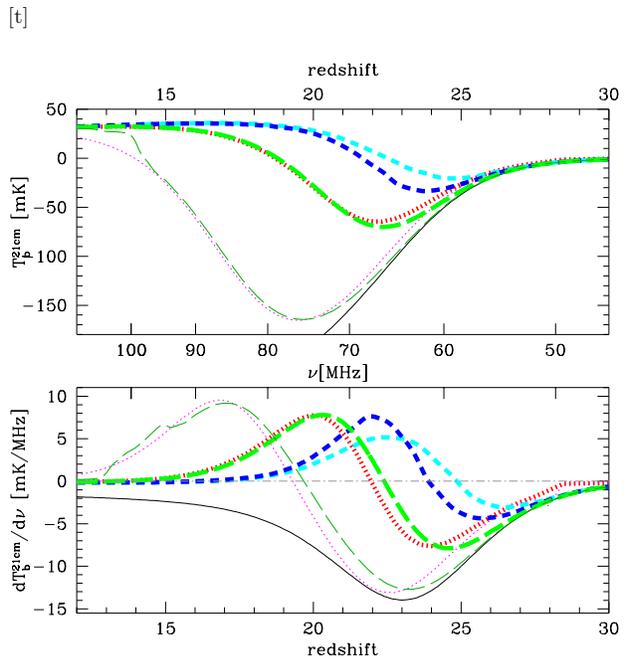}
\end{center}
\caption{\label{fig:T21} The brightness temperature of the 21~cm
  radiation (top panel), and its spectrum (bottom panel) for a medium
  heated by HMXBs alone, or with the inclusion of accreting BHs either
  in the DCBH scenario or in the PopIII formation scenario. Line types
  and colors refer to the same models as in Fig.~\ref{fig:TIGM}.}
\vspace{-1\baselineskip}
\end{figure}

After the first stars form, the thermal history of the IGM is directly imprinted in the 21\,cm brightness temperature.  One of the most important characteristics of the global, sky-averaged 21\,cm signal is when it reaches its minimum.  This turning point occurs when the IGM transitions from primarily adiabatic cooling to the epoch of heating \citep{Furlanetto06,PritchardLoeb08,Mirocha+13}, which is sensitive to the accretion history of the first black holes.  A number of experiments have been proposed, or are currently underway, to detect this signal---e.g.
the Dark Ages Radio Explorer\footnote{\url{http://lunar.colorado.edu/dare/}} (DARE; \citealt{DARE12}),
the Large-Aperture Experiment to Detect the Dark Ages\footnote{\url{https://www.cfa.harvard.edu/LEDA/}} (LEDA),
the Experiment to Detect the Global EoR Step (EDGES; \citealt{EDGES08}), and
SCI-HI (\citealt{SCIHI14}).

The upper panel of Figure \ref{fig:T21} shows 
the 21~cm brightness temperature $T_b^{21{\rm cm}}$ as a function of redshift for the same models
shown in Figure \ref{fig:TIGM}. These models use the same line styles and follow the same order as discussed in \S~\ref{sec:IGM}. Because PopIII BH models rapidly accrete gas, they begin heating the IGM as soon as they form.  The total amount of X-ray heating dominates over stellar sources (HMXBs and supernova remnants). 
As a result, our PopIII BH models increase in brightness temperature beginning at $z\la 25$,
at typically $\approx 20$ mK above the adiabatic case.
The detection of such a signature would strongly suggest either (i) stellar populations from the earliest galaxies
produced far more X-rays per unit mass formed in stars than what is observed from nearby star-forming galaxies,
or (ii) nuclear BHs were prolifically emitting X-rays as mini-AGN at these redshifts (much earlier than predicted by most DCBH scenarios).

In contrast to the PopIII models, the DCBH models do not produce sufficient X-rays to significantly heat the IGM before $z\gsim 20$.  HMXBs produced in ordinary stellar evolution  dominate the IGM heating while the 21 cm signal is in absorption. Because the gas has more time to adiabatically cool, the resulting signal is intrinsically stronger and easier to detect with a lower foreground sky temperature \citep{Furlanetto+06}.
If the 21 cm trough is detected at late times, either the PopIII BHs were heavily obscured, or the BHs that are the progenitors of SMBHs formed via direct collapse.

 At the frequencies required to detect the dark ages  ($50 - 90\,$MHz) the observations will be dominated by {\em smooth} Galactic foreground emission. The 21 cm signal contributes only about 1 part in $10^4$ to the total signal. The 21 cm signal is detectable by the variations of the brightness temperature with frequency.    The lower panel of figure~\ref{fig:T21}  shows the derivative of the brightness temperature, ${\rm d} T_b^{21\,{\rm cm}}/{\rm d}\nu$, as a function of $z$, for the same models.  

An upward trend in the 21~cm brightness temperature prior to $z\gsim 20$
strongly implies rampant PopIII seed growth, or a much higher X-ray production
from young stellar populations than seen in the local Universe.  At these frequencies, planned experiments are dominated by systematic uncertainties.  For example, DARE will have a measurement uncertainty of order 1 mK at these frequencies after 3000 h of integration. In addition, it is capable of determining when the brightness temperature reaches a minimum to $\delta z \approx 0.4$. Typical uncertainties in the minimum temperature are expected to be $\delta T_b^{21\,{\rm cm}} / T_b^{21\,{\rm cm}} \approx 15\%$ for models similar to the ones considered here \citep{Harker+12}.   

Finally, more stringent constraints on the progenitors of SMBHs may be made by combining the global 21 cm observations with other techniques.  For example, observations of the evolution of star formation at high redshift may be able to constrain the contribution of stellar sources to the IGM heating. Angular correlations in the 21 cm signal \citep{PritchardLoeb08,McQuinnOLeary12, Visbal+12} may also be able to ascertain the characteristic dark matter halo masses that dominate the IGM heating.

Both reionization and thermal feedback on baryonic structure formation
would suppress the 21~cm brightness temperature at late times.
An upward trend in the brightness temperature at $z\la 15$
would therefore strongly imply that SMBH growth began in earnest at this epoch.

\section{Summary}
\label{sec:summary}

IGM heating by the first radiation sources leaves a marked imprint on
the 21~cm radiation. Here, using both analytical arguments as well as
Monte Carlo simulations, we have argued that the primary driver of IGM
heating are the BH remnants of PopIII stars, if these are the
progenitors of the SMBHs observed in the Universe. In this scenario,
heating by these sources would dominate over the other main
constribution, i.e. HMXBs, unlike the case in which SMBHs grow from
DCBHs, which we find to be subdominant with respect to HMXBs.
 
Our work in estimating the 21 cm signature of the progenitors of SMBHs is distinct in three respects. First, the total BH mass density
in our models does not grow in proportion to the total mass density locked inside dark matter haloes \citep[as in most previous works, e.g.][]{Furlanetto06,PritchardLoeb08,Mirocha14}, but are grown individually while reflecting effects such as the Eddington accretion limit, mergers, and gravitational recoil.
Second, we explicitly treat the propagation and absorption of the X-rays averaged over the cosmological volume \citep{Mirocha14}.  Finally, our models are constrained to those that reproduce the properties of high redshift quasars \citep{TPH12, TLH13}.

The 21~cm signal, with its sensitivity to the heating history of the
Universe, hence becomes a powerful tool to probe the progenitors of SMBHs,
and their growth history.  In particular, we have shown that a trend upward
in the brightness temperature of the 21~cm radiation would indicate an abundant
production of PopIII seed BHs, and hence would lend support to a scenario
in which the growth of the SMBHs  originated from those.

\section*{Acknowledgments}
We thank the reviewer, Emanuele Ripamonti, for constructive comments that helped improve the clarity of the manuscript.

\end{document}